\documentclass[twocolumn,aps,superscriptaddress,unsortedaddress]{revtex4}
\usepackage{graphicx}
\usepackage{epsf}

\begin{document}

\title{Electronic phase diagram of high temperature copper oxide superconductors}

\author{U. Chatterjee}\author{J. Zhao}\author{D. Ai}
\affiliation{Department of Physics, University of Illinois at Chicago, Chicago, IL 60607}
\affiliation{Materials Science Division, Argonne National Laboratory, Argonne, IL 60439}
\author{S. Rosenkranz}
\affiliation{Materials Science Division, Argonne National Laboratory, Argonne, IL 60439},
\author{A. Kaminski}
\affiliation{Ames Laboratory and Department of Physics and Astronomy,  Iowa State University, Ames, IA  50011},
\author{Raffy}\author{Z. Z. Li}
\affiliation{Laboratorie de Physique des Solides, Universite Paris-Sud, 91405 Orsay Cedex, France},
\author{K. Kadowaki}
\affiliation{Institute of Materials Science, University of Tsukuba, Ibaraki 305, Japan},
\author{M. Randeria}
\affiliation{Department of Physics, The Ohio State University, Columbus, OH  43210}
\author{M. R. Norman}
\affiliation{Materials Science Division, Argonne National Laboratory, Argonne, IL 60439} 
\author{J. C. Campuzano}
\affiliation{Department of Physics, University of Illinois at Chicago, Chicago, IL 60607}
\affiliation{Materials Science Division, Argonne National Laboratory, Argonne, IL 60439}

\begin{abstract}
In order to understand the origin of high-temperature super- conductivity in copper oxides, we must understand the normal state from which it emerges. Here, we examine the evolution of the normal state electronic excitations with temperature and car- rier	concentration	in Bi$_2$Sr$_2$CaCu$_2$O$_{8+\delta}$ using angle-resolved photo- emission. In contrast to conventional superconductors, where there is a single temperature scale $T_c$ separating the normal from the superconducting state, the high-temperature superconductors exhibit two additional temperature scales. One is the pseudogap scale $T^*$, below which electronic excitations exhibit	 an energy gap. The	second is	the coherence scale	$T_{\textrm{coh}}$, below	 which sharp spectral features appear due to increased lifetime of the excitations. We find	that  $T^*$ and  $T_{\textrm{coh}}$ are strongly doping dependent and cross each other near optimal doping. Thus the highest superconducting Tc emerges from an unusual normal state that is characterized by coherent excitations with an energy gap.

\end{abstract}

\maketitle{

General features of the phase diagram of the copper oxide superconductors have been known for some time. The superconducting transition temperature $T_c$ has a dome-like shape in the doping-temperature plane with a maximum near a doping $\delta \sim$ 0.167 electrons per Cu atom. While in conventional metals the electronic excitations for $T > T_c$ are (i) gapless and (ii) sharply defined at the Fermi surface\cite{JRS}, the cuprates violate at least one of these conditions over much of their phase diagram. These deviations from conventional metallic behavior are most easily described in terms of two energy scales $T^*$\cite{DINGPG,LOESERPG} and $T_{\textrm{coh}}$\cite{ADAMCOHERENCE}, which correspond to criteria (i) and (ii), respectively.

To address the role of these energy scales in defining the phase diagram, we concentrate on spectra where the superconducting energy gap is largest, the antinode ($(\pi,0)\to(\pi\pi)$ Fermi crossing),  where the spectral changes with doping and temperature are most pronounced (See SI for further details). Spectral changes at the node have been previously studied by Valla et \textit{al.}\cite{VALLASCATRATE} and such spectra remain gapless for all doping values\cite{UTPALNODAL}. In Fig. 1 we show spectra at fixed temperature as a function of doping. Data points are indicated in Fig. 1{\it A} (See SI for experimental conditions and sample details). Initially, we show spectra at fixed momenta as a function of energy (energy distribution curves, or EDCs) that have been symmetrised\cite{DESTRUCTION} about the Fermi energy to remove the effects of the Fermi function. Later, we show that equivalent results are obtained from division of the EDCs by a resolution-broadened Fermi function.

The spectra at the antinode at the highest temperature ($\sim$300K) in Fig. 1{\it D} show two remarkable features: they are extremely broad in energy, exceeding any expected thermal broadening, and their lineshapes, well described by a lorentzian, are independent of doping. The large spectral widths indicate electronic excitations that cannot be characterized by a well-defined energy, implying that the electrons are strongly interacting. 
 
 \begin{figure}
\centerline{\includegraphics[width=.5\textwidth]{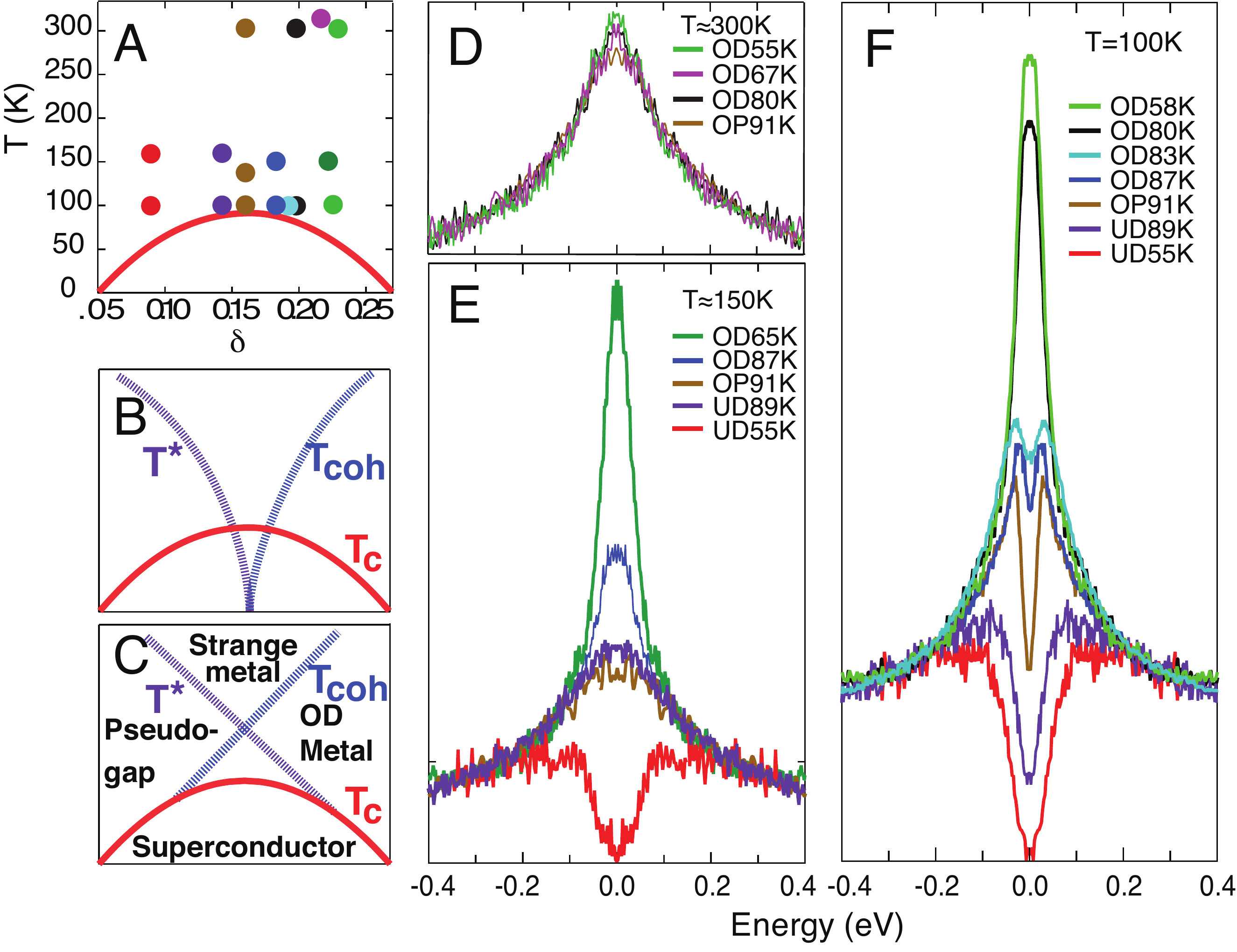}}
\caption{
Spectra at constant temperature as a function of doping. ({\it A}) Dots indicate the temperature and doping values of the spectra of the same color plotted in ({\it D-F}). ({\it B}) Schematic phase diagram for a quantum critical point near optimal doping. ({\it C}) Schematic phase diagram for a doped Mott insulator. ({\it D}) Spectra at $T\sim 300$K for several samples measured  at the antinode, where the $d$-wave superconducting gap below $T_c$ is largest.  The spectra are normalized to high binding energy and symmetrised in energy to eliminate the Fermi function. The doping values are indicated by the top row of dots in ({\it A}). ({\it E}) Same as in ({\it D}), but at $T \sim 150$K, with the dopings indicated by the middle row of dots in ({\it A}). ({\it F}) Same as in ({\it D}), but at $T=100$K, with the dopings indicated by the  bottom row of dots in ({\it A}).
}\label{fig1}
\end{figure}

The incoherent behavior of the spectra at 300K is consistent with the Ôstrange metalÕ regime in two model phase diagrams popular in the literature, shown schematically in Figs. 1{\it B} and 1{\it C}.  If Fig. 1{\it B} applies, there would be strong evidence for a single quantum critical point near optimal doping which dominates the behavior to high temperatures\cite{VARMAPGTHEORY, SUBIRWHEREQCP}. $T^*$ would be the transition temperature for a competing order, with $T_{\textrm{coh}}$ its 'mirror' corresponding to where Fermi liquid behavior sets in. The non-Fermi liquid behavior in the Ôstrange metalÕ phase above both scales would then arise from fluctuations in the quantum critical region. These same fluctuations presumably mediate superconducting pairing. On the other hand, if Fig. 1{\it C} applies, the phase diagram would arise from strong correlation theories based on doped Mott insulators\cite{PWASCIENCE,FUKUYAMA, GABI,PALEE}. The $T^*$ line is where spin excitations become gapped, whereas $T_{\textit{coh}}$ is the temperature below which doped carriers become coherent. Superconductivity emerges below both scales, where spin and charge excitations become gapped and coherent. Which of these two phase diagrams is the appropriate one has critical implications for our understanding of the cuprates. To study this, we reduce $T$.
At $\sim$150K, the spectra show marked changes with doping, and three regions can be identified in Fig. 1{\it E}. At low $\delta$, the spectrum (red curve) remains broad as in Fig. 1{\it D}, but now a spectral gap is present - the pseudogap. This results in a reduction of the low-energy spectral weight as probed by various experiments\cite{TIMUSKSTATT}. On increasing $\delta$, the spectral gap becomes less pronounced, and disappears just below optimal doping (purple and brown curves), where the spectra now resemble those in Fig. 1{\it D}. Increasing $\delta$ beyond 0.17, the spectra exhibit a sharp peak centered at zero energy ($E_F$) (blue and green curves). It can be seen in Fig. 1{\it E} that the sharper portion of the latter two spectra rises above the lorentzian part of the spectrum delineated by the purple curve. These sharp spectra are now similar to what one would expect for a more conventional metal\cite{JRS,VALLACOHERENCE}. The doping dependences near 150K are again consistent with either Fig. 1{\it B} or 1{\it C}.
 
A completely different behavior emerges at a lower temperature, 100K (Fig. 1{\it F}). The pseudogap with no sharp peaks is still present for low $\delta$ (red and purple curves). But near optimal doping, the spectra change, now exhibiting sharp peaks separated by an energy gap (brown, blue and light blue curves). These sharp peaks indicate that the lifetimes of excitations have increased dramatically, in contrast to the spectra at 150K for optimal doping (brown curve in Fig. 1{\it E}). For still higher $\delta$, a single sharp peak centered at $E_F$ appears (green and black curves). Notice that all the spectral changes are limited to an energy scale of less than 200 meV; outside this energy range, the spectra follow the same broad lorentzian shape as in Fig. 1{\it D}. 

\begin{figure}
\begin{center}
\centerline{\includegraphics[width=.5\textwidth]{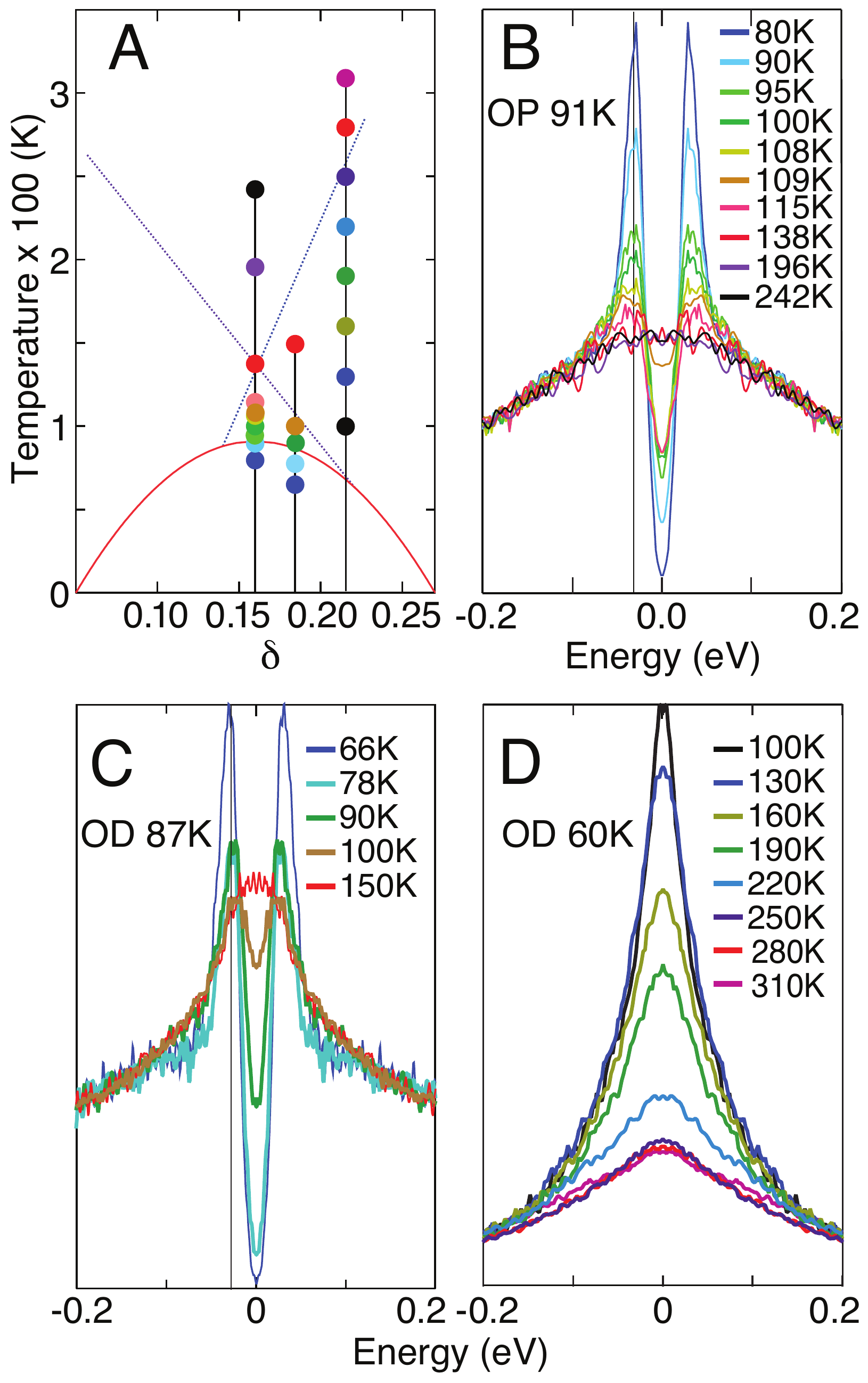}}
\caption{
Spectra at constant doping as a function of temperature. ({\it A}) Dots indicate the temperature and doping values of the spectra of the same color plotted in ({\it B - D}). ({\it B}) Symmetrised antinodal spectra for two optimally doped samples ($\delta = 0.16$). The temperature values are indicated by the left row of dots in ({\it A}). ({\it C}) Same as in ({\it B}), but for a doping of $\delta = 0.183$, with the temperatures indicated by the middle row of dots in ({\it A}). ({\it D}) Same as in ({\it B}), but for a doping of $\delta = 0.224$, with the temperatures indicated by the right row of dots in ({\it A}). Gray lines in ({\it B}) and ({\it C}) mark the location of the gap.
}\label{fig2}
\end{center}
\end{figure}
 
Fig. 1{\it F} demonstrates that the spectral gap and coherence (manifested by sharp spectral peaks) coexist in the normal state near optimal doping, implying that the $T^*$ and $T_{\textit{coh}}$ lines cross each other, as in Fig. 1{\it C}. To further illustrate this crossing, we plot spectra at fixed doping as a function of temperature, with the various data points indicated in Fig. 2{\it A}. Fig. 2{\it B} shows spectra for two optimally doped samples. At $T = 90$K (light blue curve) the sample is just emerging from the superconducting state. Increasing $T$, the sharp peaks at the edge of the gap decrease in intensity, while the gap magnitude remains constant. Finally, for $T \ge 115$K (pink curve), the sharp peaks disappear, while the spectral gap remains. This indicates that the $T_{\textrm{coh}}$ line has been crossed, but not the $T^*$ line. For $T \ge 138$K (red curve), the spectral gap has completely filled in, and the spectra have regained the broad, temperature-independent lineshape characteristic of the Ôstrange metalÕ phase of Fig. 1{\it D}. 
 
In contrast, upon increasing the doping, the crossing of the pseudogap and coherence temperatures are reversed, as illustrated in Fig. 2{\it C}. Starting in the superconducting state at 66K (blue curve), one can see the same features as in Fig. 2{\it B}, but now the spectral gap is smaller. Once $T_c$ is crossed at 90K, the spectral gap and sharp peaks persist (green and brown curves).  But at higher $T$, the gap disappears, and we are left with a relatively sharp peak at $E_F$ (red curve), in contrast to Fig. 2{\it B}. For higher temperatures, the peaks broaden as in Fig. 1{\it D}. If the doping is now increased even further (Fig. 2{\it D}), a spectral gap is no longer observed at any $T > T_c$.  In this highly overdoped region, the superconducting transition is similar to that of conventional superconductors\cite{JRS}, as the spectral gap closes very near $T_c$. The peak at $E_F$ initially remains sharp, but at high enough temperatures, the Ôstrange metalÕ returns (purple, red, and violet curves).

\begin{figure}
\centerline{\includegraphics[width=.5\textwidth]{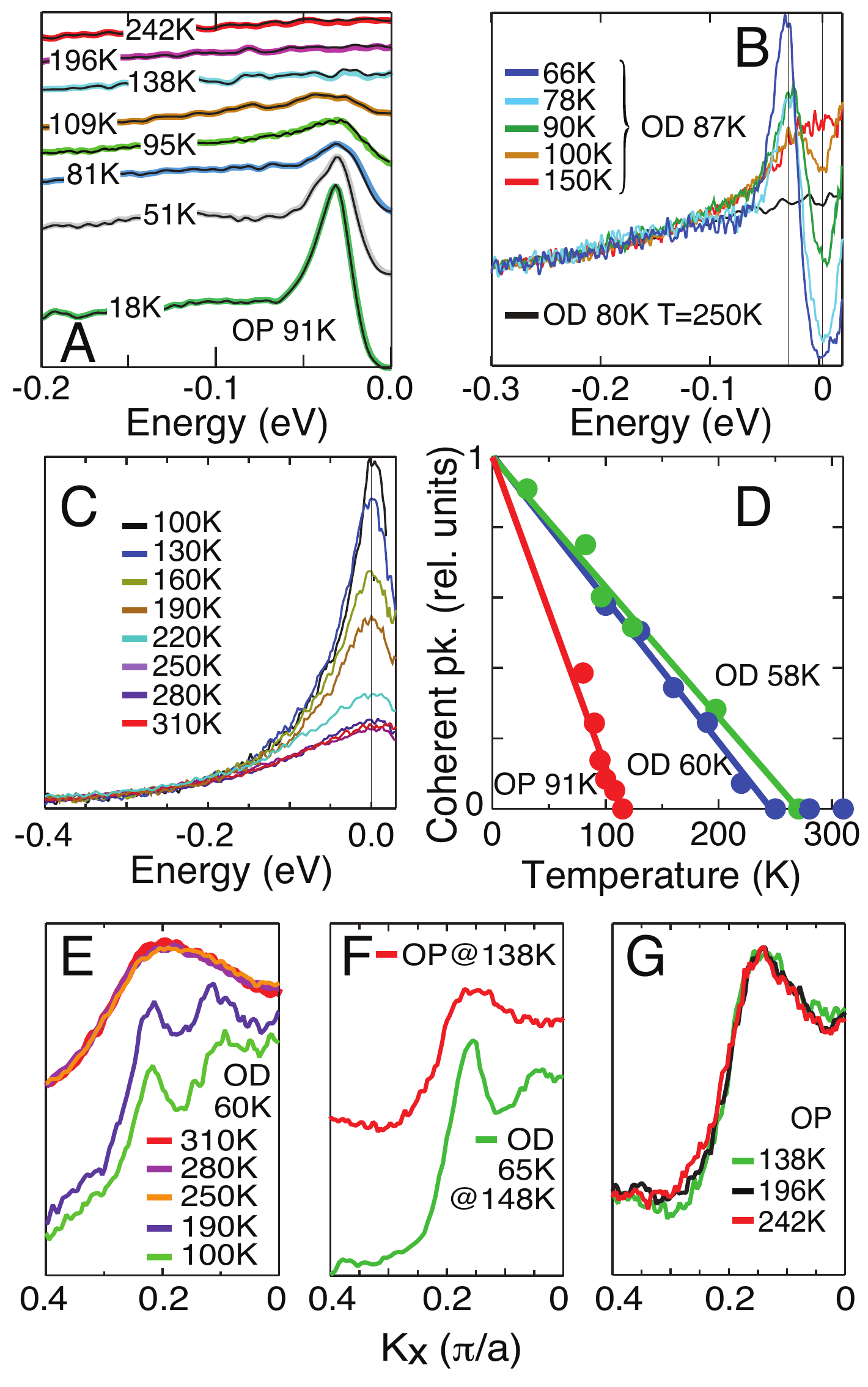}}
\caption{
Fermi function divided spectra. ({\it A}) Antinodal spectra for two optimally doped samples, $\delta = 0.16$, showing sharp peaks with an energy gap (green curves) below $T^*$, but broad gapless spectra (purple curve) above $T^*$. Colored lines show Fermi function-divided data, with symmetrised data superimposed as sharp black lines. ({\it B}) Spectra for an overdoped sample, $\delta = 0.183$, showing that unlike in ({\it A}), the spectral gap is lost above 100K, while the sharp peak persists to higher temperature. ({\it C}) Data for an overdoped $\delta= 0.224$ sample. The sharp spectral peak decreases in intensity with increasing temperature. By $T = 250$K, the spectral lineshape is broad and temperature independent. ({\it D}) Linearly decreasing intensity of the sharp spectral peak relative to the broad lorentzian with increasing $T$ for three values of $\delta$. $T_{\textrm{coh}}$ is where this intensity reaches zero. ({\it E}) MDCs for an overdoped $\delta = 0.224$ sample, showing that a qualitative change in spectral shape occurs near $T_{\textrm{coh}}$. ({\it F}) Comparison of the MDC of an OP doped sample, to that of an OD sample at a similar $T$.({\it G}) $T$-independence of the spectral shape for an OP sample above $T_{\textrm{coh}}$.
}\label{fig3}
\end{figure}
 
In Fig. 3 we show that dividing the EDCs by a resolution-broadened Fermi function\cite{MATSUIBOGOLIUBOV} gives equivalent results to symmetrising them.  To quantitatively determine the $T^*$ line, we note that it is easily identified by where the spectral gap disappears\cite{AMITSCALING}. For $T_{\textrm{coh}}$ we need to identify where the sharp peak disappears. We find that we can model the broad, incoherent part of the spectrum with a lorentzian centered at $E_F$, and the sharp, coherent piece with a gaussian (for details, see SI). In Fig. 3{\it D} we plot the height of the sharp component of the spectra above that of the constant lorentzian. One clearly sees a linear decrease with $T$, from which we determine $T_{\textrm{coh}}$. $T_{\textrm{coh}}$ can also be observed in plots of the ARPES signal as a function of momentum for a fixed energy, the momentum distribution curves (MDCs) shown in Figs. 3{\it E, F, G}. In Fig. 3{\it E} we show that a significant change in width occurs upon crossing $T_{\textrm{coh}}$, which clearly indicates that this is not a simple temperature broadening effect. The spectra remain relatively unchanged both below and above $T_{\textrm{coh}}$, with significant changes limited to temperatures close to $T_{\textrm{coh}}$. Furthermore, $T_{\textrm{coh}}$ is strongly doping dependent. In Fig. 3{\it F} we show spectra at similar $T$ for an optimally doped sample, and an overdoped one with $T_c = 65$K, showing that the spectral widths depend on the region of the phase diagram, and not simply the temperature. This is emphasized in Fig. 3{\it G}, where no spectral changes are observed in the strange metal region over a wide range in temperature.

\begin{figure}
\centerline{\includegraphics[width=.5\textwidth]{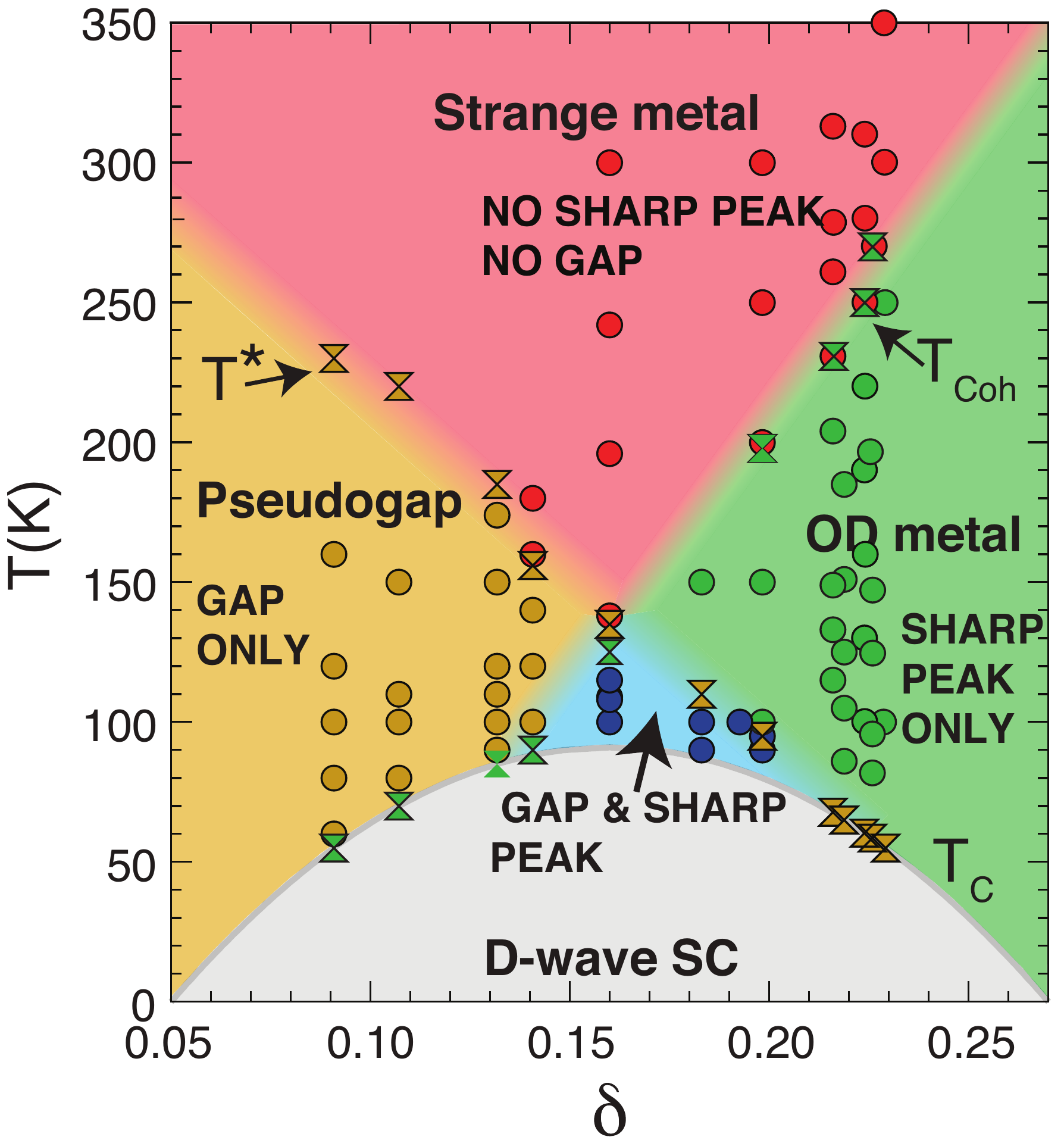}}
\caption{
Electronic phase diagram of Bi$_2$Sr$_2$CaCu$_2$O$_{8 + \delta}$ versus hole doping, $\delta$.  Brown dots indicate incoherent gapped spectra, blue points coherent gapped spectra, green dots coherent gapless spectra, and red dots incoherent gapless spectra.  The brown double triangles denote $T^*$, and the green double triangles $T_{\textrm{coh}}$. $T_c$ denotes the superconducting transition temperature.
}\label{fig4}
\end{figure}
      
The phase diagram shown in Fig. 4 summarizes our results. The solid dots are based on the antinodal spectra and are color coded to correspond to the four different regions in the normal state phase diagram. These correspond to antinodal spectra that are: (1) incoherent and gapped (brown dots), in the underdoped pseudogap region, (2) incoherent and gapless (red dots), in the high temperature Ôstrange metalÕ, (3) coherent and gapless (green dots), in the overdoped metal, and finally (4) coherent and gapped (blue dots), in the triangular region above optimal doping formed as a result of the crossing of $T^*$ and $T_{\textrm{coh}}$. In addition, we also plot $T^*$ and $T_{\textrm{coh}}$ as defined above by double triangles. We emphasize that below $T_c$, we find coherent and gapped antinodal spectra for all doping values, even for very underdoped samples\cite{UTPALNODAL}.  

An earlier ARPES experiment showed the appearance of dichroism below a temperature equivalent to the $T^*$ measured here\cite{ADAMTRS}, as did subsequent neutron scattering experiments that detected intra-unit cell magnetic order\cite{BOURGES,GREVEN}, both of which identify $T^*$ as a phase transition. However, the present experiments do not measure an order parameter. Moreover, it is not clear that the large energy gap is due to magnetism. We therefore limit ourselves to calling $T^*$ a 'temperature scale'.
    
Although heat capacity\cite{TALLONLORAMREALPD} and more recent transport studies\cite{HUSSEY} have suggested Fig. 1{\it B}, transport represents a single static (dc) quantity. On the other hand, photoemission being an energy and momentum resolved probe, allows one to uniquely separate the influence of coherence, a spectral gap, and their momentum dependence. In further support of Fig. 4, we note that the high-doping side of the blue triangle near optimal doping, characterized by gapped and coherent spectra above $T_c$, has also been inferred from the $T$ dependence of scanning tunneling spectra \cite{GOMES}. To our knowledge, however, the full triangular region has not been identified before. Although at first sight this triangular region seems similar to the region where diamagnetism is observed above 
$T_c$\cite{ONG}, the latter
has a larger extent over the phase diagram than the former.  This is not a surprise, since we are measuring single particle coherence, whereas the diamagnetism is
a measure of superconducting fluctuations.

Our experimental finding that the intersect of the two temperature scales is not consistent with a single quantum critical point near optimal doping, although more complicated quantum critical scenarios cannot be ruled out. For instance, quantum critical points exist at the ends of the dome\cite{BROUNQCPDOMEEND,HETELQCPDOMEEND}. In our data, Fig. 4, superconductivity only emerges below both $T^*$ and $T_{\textrm{coh}}$. And, optimal superconductivity emerges from a coherent, gapped, normal state. Hence, our results are more naturally consistent with theories of superconductivity for doped Mott insulators, as illustrated in Fig. 1{\it C}. We believe these results represent an important step forward in solving the highly challenging problem of high temperature superconductivity.
\\U.C. and J.C.C. designed research; U.C., D.A., J.Z., S.R., and A.K. performed research; H.R., Z.Z.L. and K.K. grew the samples. U.C., M.R.N. M.R., and J.C.C. analyzed data and wrote the paper.
\\$^1$To whom correspondence should be addressed. E-mail: jcc@uic.edu.

\begin{acknowledgments}
This work was supported by the National Science Foundation under grant DMR-0606255 (J.C.C.), and NSF-DMR 0706203 (M.R.). Work performed at the Synchrotron Radiation Center, University of Wisconsin (Award No. DMR-0537588). The work at Argonne National Laboratory was supported by UChicago Argonne, LLC, Operator of Argonne National Laboratory. Argonne, a U.S. Department of Energy, Office of Science laboratory is operated under Contract No. DE-AC02-06CH11357 (S.R., A.K., M.R.N, and J.C.C.).
\end{acknowledgments}

\end{document}